\documentclass{amsart}
\usepackage{graphicx,natbib,subfigure}
\usepackage[squaren]{SIunits}

\begin{document}

\title{Stress Orientation Confidence Intervals from Focal Mechanism Inversion}

\author{S. A. Revets} 

\address{School of Earth \& Environment\\
University of Western Australia\\
35 Stirling Highway M004\\
Crawley WA6008, Australia}

\email{stefan.revets@uwa.edu.au}

\thanks{I thank the Australian Research Council, the Australian National
University, Woodside Petroleum and Geoscience Australia for supporting
this research through ARC Linkage Grant LP0560955.}

\begin{abstract}
  The determination of confidence intervals of stress orientation is a
  crucial element in the discussion of homogeneity or heterogeneity of
  the stress field under study. The error estimates provided by the
  grid search method Focal Mechanism Stress Inversion of
  \citet{gephart-forsyth84:improved} have been shown to be too wide
  but the reasons for this failure have escaped elucidation.

  Through the use of directional statistics and synthetic focal
  mechanisms, I show that the grid search methodology does yield
  appropriate uncertainty estimates. The direct perturbation of the
  synthetic focal mechanisms introduces bias which leads to confidence
  intervals which become increasingly too wide as the amount of
  perturbation increases. The synthetic data also show at what point
  the method fails to overcome this bias and when confidence intervals
  will be too wide. The indirect perturbation of the focal mechanisms
  by perturbing the generating deviatoric stress tensor generates
  synthetic data devoid of bias. Inversion of these data sets yields
  correct confidence intervals.

  The Focal Mechanism Stress Inversion method is vindicated as a
  highly effective method, and with the use of appropriate directional
  statistics, its results can be assessed and homogeneity or
  heterogeneity of the stress field can be discussed with confidence.
\end{abstract}

\maketitle

\section{Introduction}
\citet{gephart-forsyth84:improved} proposed a grid search method of
inverting focal mechanisms to obtain the stress tensor (focal
mechanisms stress inversion, henceforward FMSI), in which stress field
parameters are tried systematically against the focal mechanism
orientations and the misfit calculated. They defined the misfit as the
sum of the minimum angle needed to bring the slip direction of each
focal mechanism into line with the resolved shear stress on the fault
plane. Both planes of each focal mechanism are tried and the smallest
misfit serves to differentiate fault plane from auxiliary plane.  They
incorporated the one-norm measure as misfit criterion, adopting the
methodology proposed by \citet{parker-mcnutt80:one-norm}.  Different
techniques for such inversions had been proposed already, and
\citet{gephart90:stress} discussed their relative merits, particularly
in the context of focal mechanisms and the problems associated with
the presence of two nodal planes.

\citet{michael87:stress} proposed a different method, relying on a
linearisation and bootstrapping to invert focal mechanisms for stress
tensor calculation. He drew attention to some differences in the size
of the confidence intervals obtained by these two different methods.

It became quickly apparent that the discrepancies in confidence
intervals had major repercussions on the study of actual stress
fields. Different methods led different groups of researchers to
different conclusions regarding the spatial and temporal variation in
the stress regime in Southern California. This unsatisfactory state of
affairs led \citet{hardebeck-hauksson01:stress} to test thoroughly the
different methodologies using synthetic data.  They showed that the
confidence intervals estimated by FMSI were much too large, but they
did not succeed in determining the reasons for such overestimates.

Here, I attempt to uncover the nature of the confidence intervals
estimated by FMSI by investigating the
\citet{hardebeck-hauksson01:stress} analysis. Initial attempts to
resolve the problem of these too wide confidence intervals brought to
light some inaccuracies in the derivation and application of the
one-norm measure as proposed by \citet{parker-mcnutt80:one-norm} and
applied by \citet{gephart-forsyth84:improved} and
\citet{hardebeck-hauksson01:stress} \citep{revets09:one-norm}. These
corrections proved insufficient to lessen significantly the width of
the confidence intervals (Hardebeck, pers.\ comm.).

This left two avenues of investigation: the nature of the statistics
used to calculate the confidence intervals, and the methodology of
calculating the synthetic data used to test the method.

\section{Directional statistics}
\citet{fisher53:dispersion} pointed out that the initial development
of the theory of errors by Gauss aimed to help astronomers and
surveyors with their accurate angular measurements. Because of the
accuracy of their measurements, the linear approximation to which
Gauss resorted was both appropriate and effective. However, when the
errors become substantial, the linear approximation is no longer valid
and the topological framework has to be taken into account.

The spherical mean direction $R$ and variance $S$ are defined as
\begin{equation}
  \label{eq:spherical-mean}
  R^2 = \big(\sum_{i=1}^N l_i\big)^2 + \big(\sum_{i=1}^N m_i\big)^2 + \big(\sum_{i=1}^N n_i\big)^2
\end{equation}
and
\begin{equation}
  \label{eq:spherical-variance}
  S = (N - R)/N
\end{equation}
where $l_i$, $m_i$, $n_i$ are the direction cosines of the angular
measurements \citep{mardia72:directional}.

The spherical equivalent of the normal distribution is the Fisher
distribution \citep{fisher53:dispersion}, defined by
\begin{equation}
  \label{eq:fisher-distribution}
  df = \frac{\kappa}{2 \sinh \kappa} e^{\kappa \cos \theta}\sin \theta d\theta
\end{equation}
in which $\kappa$ is a measure of concentration. When $\kappa$
approaches zero, the distribution becomes uniform over the entire
sphere whereas for $\kappa$ large, the distribution is confined to a
small region of the sphere around the maximum. In the latter case, the
distribution comes close to a two-dimensional (isotropic) normal
distribution where $\kappa$ plays the role of the inverse of the
variance.

The maximum likelihood estimate of $\kappa$ for the distribution on a
sphere is for larger values of $\kappa$
\begin{equation}
  \label{eq:kappa-bessel}
  \bar{R} = \frac{I_{1.5}(\kappa)}{I_{0.5}(\kappa)}
\end{equation}
where $I_n$ is the modified Bessel function of the first kind and the
n-th order \citep{watson44:bessel} and $\bar{R}$ the mean of $R$
\citep{watson-williams56:construction,mardia72:directional}. For large
values of $\kappa$, the most likely value is given by
\begin{equation}
  \label{eq:kappa-R}
  \kappa = \frac{N - 1}{N - R}
\end{equation}
\citep{fisher53:dispersion,watson60:more}.

Confidence intervals for $\kappa$ can be determined from the relation
\begin{equation}
  \label{eq:kappa-chi2}
  2 \kappa (N - R) = \chi^2_{2N-2}
\end{equation}
\citep{watson-williams56:construction,mardia72:directional}.

\section{Application to stress tensor inversion}
To calculate and test the confidence intervals of stress tensors
inverted from focal mechanisms, I adapted and modified the method used
by \citet{hardebeck-hauksson01:stress}, taking particular care to
adhere to the strictures and requirements of directional statistics.

\subsection{Generating synthetic focal mechanisms}
Sets of focal mechanisms are made up from spherically randomly
oriented fault planes with slip directions determined by a chosen
stress tensor. The azimuth of the fault planes is chosen from
uniformly distributed random numbers in the $[0, 2 \pi]$ interval,
while the dip is randomly taken out of the $[0, 1]$ interval through
\begin{equation}
  \label{eq:random-planes}
  d_i \in \arccos [0, 1]
\end{equation}
to avoid the (spherical) bias that the direct selection from the $[0,
\pi/2]$ interval would cause. Sets contain either 20, 50 or 100 fault
planes.

The direction of slip on the fault plane given the stress tensor
follows from the formalism of \citet{ramsey-lisle83:techniques}
\begin{equation}
  \label{eq:slip-direction}
  l(m^2 \phi + n^2) : m(n^2 (1 - \phi) - l^2 \phi) : -n(l^2 + m^2 (1 - \phi))
\end{equation}
where $l$, $m$, $n$, are the direction cosines of the fault plane in
the stress tensor reference frame and $\phi$ is the stress shape ratio
\begin{equation}
  \label{eq:stress_shape}
  \phi = \frac{\sigma_1 - \sigma_2}{\sigma_1 - \sigma_3}
\end{equation}
This direction of slip, as the normal to the auxiliary plane,
completes the definition of the focal mechanism.

\subsection{Generating error}
\begin{figure}
  \includegraphics[width=120mm,keepaspectratio]{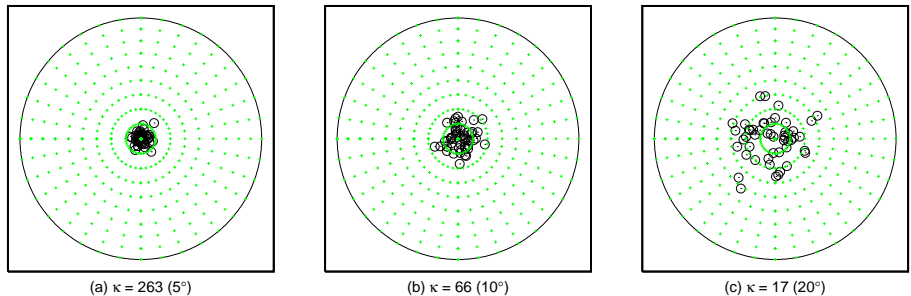}
  \caption{Polar Schmidt plots of examples of Fisher distributed
    errors, as used to perturb the generating stress tensor. The
    crosses are 10\degree\ apart}
  \label{fig:fisher-error}
\end{figure}

I used a Fisher distribution to generate random directions and angles
to perturb orientations \citep[p. 231]{mardia72:directional}, and used
quaternions to carry out the required rotations.  Figure \ref{fig:fisher-error}
shows an example of such Fisher distributed error angles and
orientations plotted on a polar Schmidt net.  Values of $\kappa$ for
the Fisher distribution which correspond to perturbations of 1\degree,
5\degree, 10\degree, 15\degree\ and 20\degree\ can be obtained through
interpolation from equation \ref{eq:kappa-bessel}.

There are two ways in which to incorporate errors in the synthetic
data set. There is the natural way of perturbing each focal mechanism
directly. It is also possible to perturb the generating stress tensor
before calculating the slip direction on each fault plane in the data
set. The assumptions behind these two ways are very different, and I
discuss their implications, statistical as well as physical, following
the simulation results.

\subsection{Data inversion and statistics}
Each synthetic data set is then processed by FMSI, which inverts the
focal mechanisms to obtain the stress tensor. Thanks to FMSIETAB, part
of the FMSI suite of programs, it is possible to generate a list of
angular deviation between each fault plane and any given stress
tensor. Such lists, comparing the deviations between focal mechanisms
with the generating stress tensor as well as with the inverted stress
tensor allow the calculation of the respective spherical mean $R$ and
the concentration parameter $\kappa$, using equations
\ref{eq:spherical-mean} and \ref{eq:kappa-R}.  Each combination of $N$
fault planes with the different amounts of perturbation is replicated
50 times.

\citet{hardebeck-hauksson01:stress} proposed that for appropriate
confidence intervals the correct stress tensor should fall within the
P confidence region for an approximate proportion P of all the data
sets. A plot of (sorted) probabilities versus proportion should be a
diagonal. If the confidence intervals are too wide, the plot will show
a curve above the diagonal and too narrow confidence intervals will
result in a curve below the diagonal. These graphs are an equivalent
of P-P plots
\citep{wilk-gnanadesikan68:probability,holmgren95:pp}. P-P plots are
scatter plots of $F_1(q_i), F_2(q_i)$, where $F_1(q_i)$ is obtained by
applying
\begin{equation}
  \label{eq:pp-plot}
F_1(F_2^{-1}(p_i))
\end{equation}
to the two empirical cumulative density functions ($F$) of the two
data sets being compared. Here, I show P-P plots which compare the
cumulative density function of the misfit against the inverted stress
tensor with the cumulative density function of the misfit against the
generating stress tensor.

\citet{gephart90:stress} proposed a modification of the method
proposed by \citet{zizicas55:representation} to represent the
orientation of a fault plane with respect to the principal stress
directions. Gephart's unscaled Mohr Sphere uses the stress components
$\sigma$, $\tau_S$ and $\tau_b$ as axes.

The maximum shear stress and mean normal stress are respectively
\begin{equation}
  \tau_m = \frac{\sigma_1 - \sigma_3}{2}, \quad \sigma_m = \frac{\sigma_1 + \sigma_3}{2}
\end{equation}
The absolute values of the stress tensor are inaccessible, but the
relative magnitudes of the principal stress components can be
calculated. Their relation is given by
\begin{equation}
  \phi = \frac{\sigma_2 - \sigma_1}{\sigma_3 - \sigma_1}
\end{equation}
The three normalized stress components acting on a fault plane can be
described as
\begin{equation}
\label{eq:stress_components}
  \sigma = \frac{\sigma'_{11} - \sigma_m}{\tau_m}, \quad
  \tau_b = \frac{\sigma'_{12}}{\tau_m}, \quad
  \tau_s = \frac{\sigma'_{13}}{\tau_m}
\end{equation}
and are completely determined by the dimensionless quantities $\phi$ and
$\beta_{ij}$, where $\beta_{ij}$ is determined from
\begin{equation}
  \label{eq:rotation_components}
  \sigma'_{ij} = \sigma_{kl} \beta_{ik} \beta{jl}
\end{equation}
which relates the stress tensor components between the coordinate
systems of the fault plane and of the regional stress tensor.  The
shear stress will match the slip direction on a plane when
$\sigma'_{12}$ is zero \citep{gephart90:stress}.

Mohr Sphere plots of fault planes are often illuminating and are of
great assistance with the interpretation of the inversions. This is
also the case with the differently treated synthetic data sets, and I
include a number of such Mohr Sphere plots for some of the data used.

\section{Results}
\subsection{P-P Plots}
\begin{figure}
\subfigure[Focal mechanism perturbation]{
  \includegraphics[width=60mm,keepaspectratio]{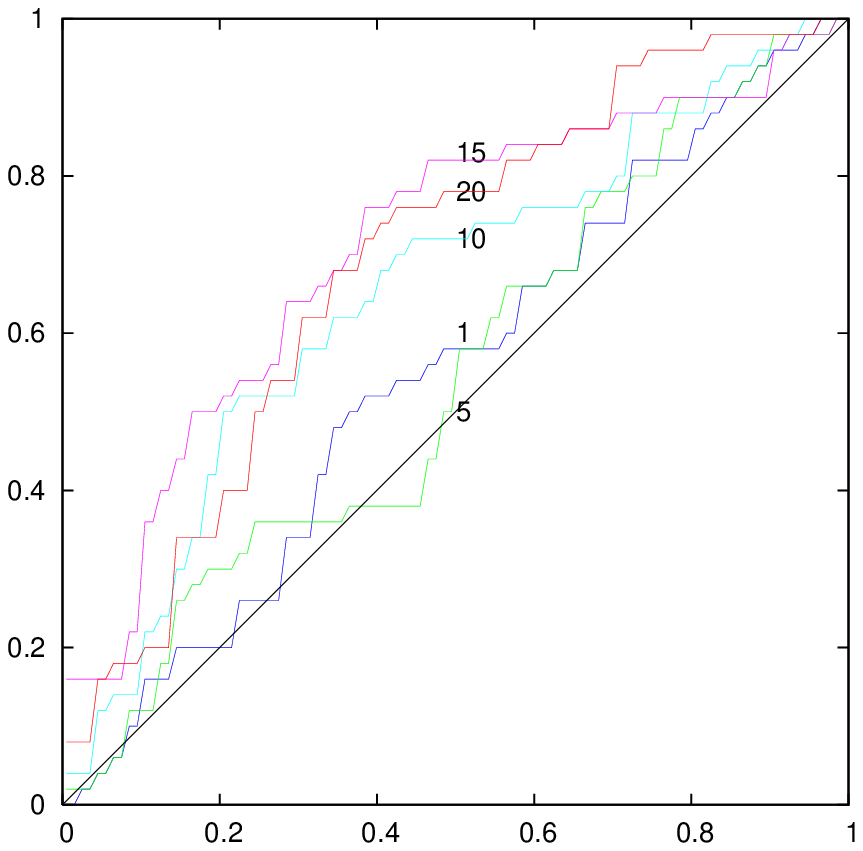}}
\subfigure[Stress tensor perturbation]{
  \includegraphics[width=60mm,keepaspectratio]{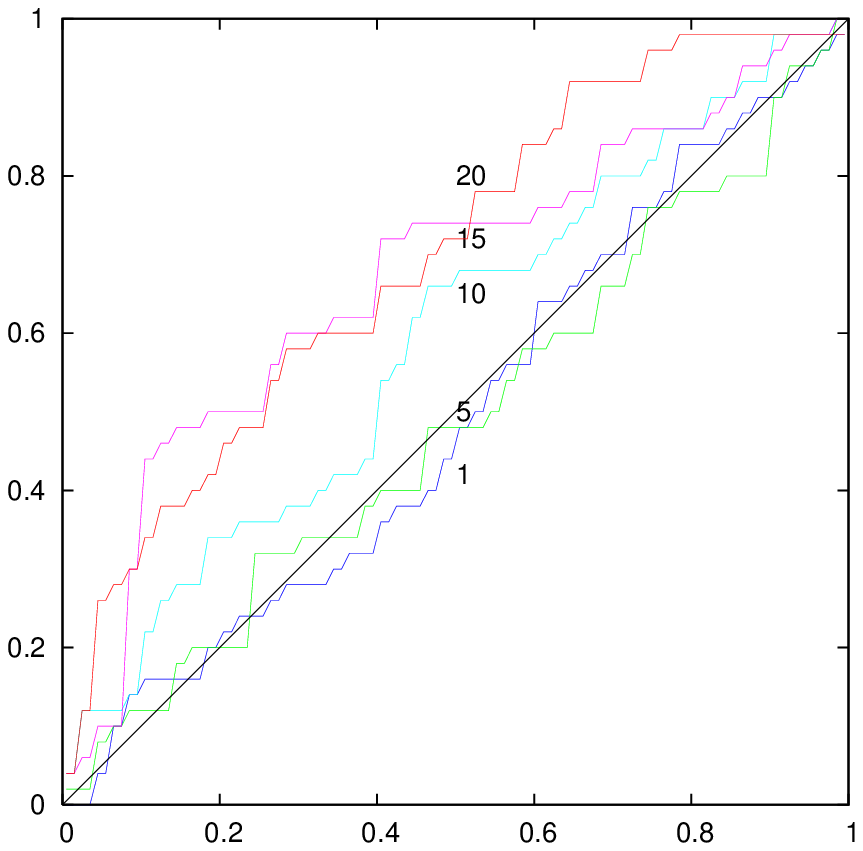}}
\caption{P--P plot of $\kappa$ for $N$=20, with the generating tensor
  values along the x-axis and the inverted tensor values along the
  y-axis. The amount of perturbation in degrees is shown on each
  individual curve.}
  \label{fig:n20}
\end{figure}

\begin{figure}
\subfigure[Focal mechanism perturbation]{
  \includegraphics[width=60mm,keepaspectratio]{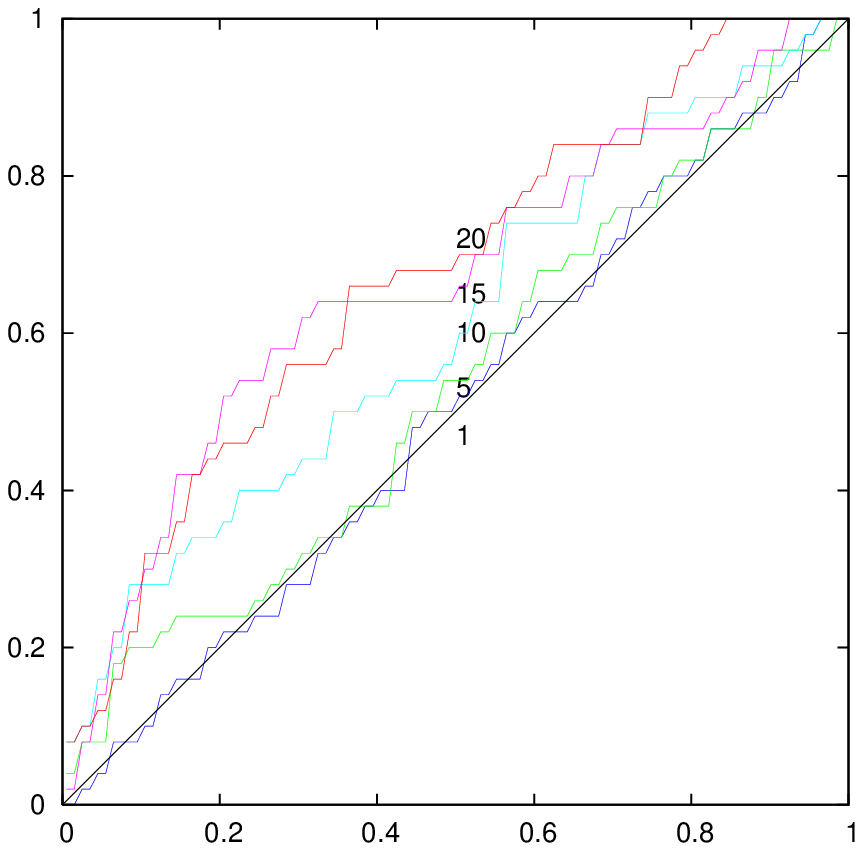}}
\subfigure[Stress tensor perturbation]{
  \includegraphics[width=60mm,keepaspectratio]{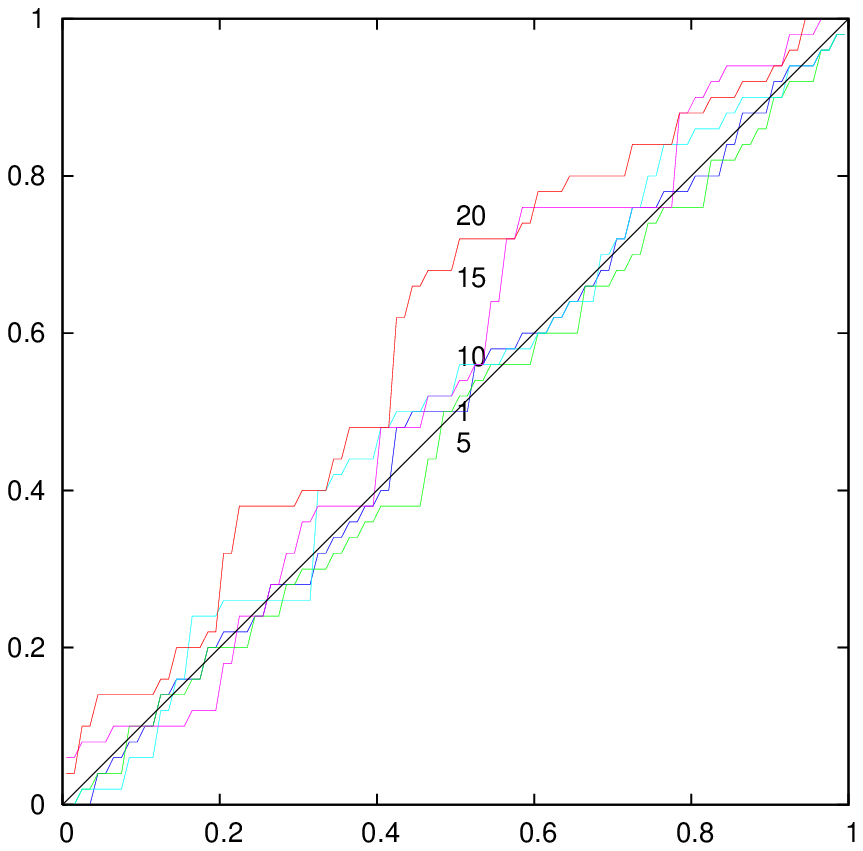}}
\caption{P--P plot of $\kappa$ for $N$=50, with the generating tensor
  values along the x-axis and the inverted tensor values along the
  y-axis. The amount of perturbation in degrees is shown on each
  individual curve.}
  \label{fig:n50}
\end{figure}

\begin{figure}
\subfigure[Focal mechanism perturbation]{
  \includegraphics[width=60mm,keepaspectratio]{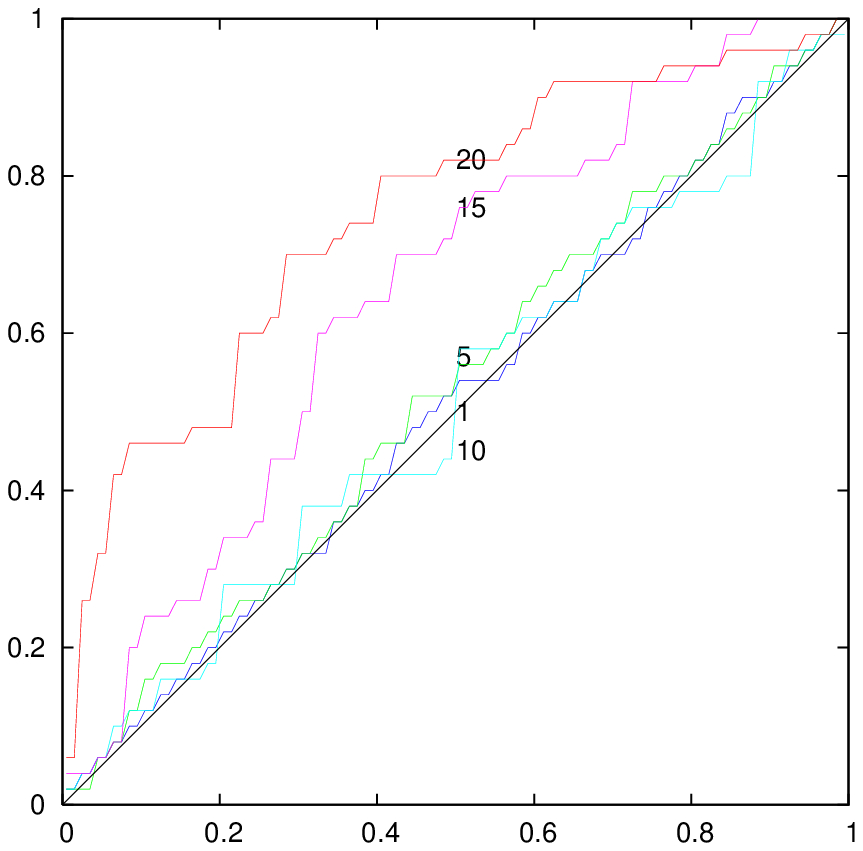}}
\subfigure[Stress tensor perturbation]{
  \includegraphics[width=60mm,keepaspectratio]{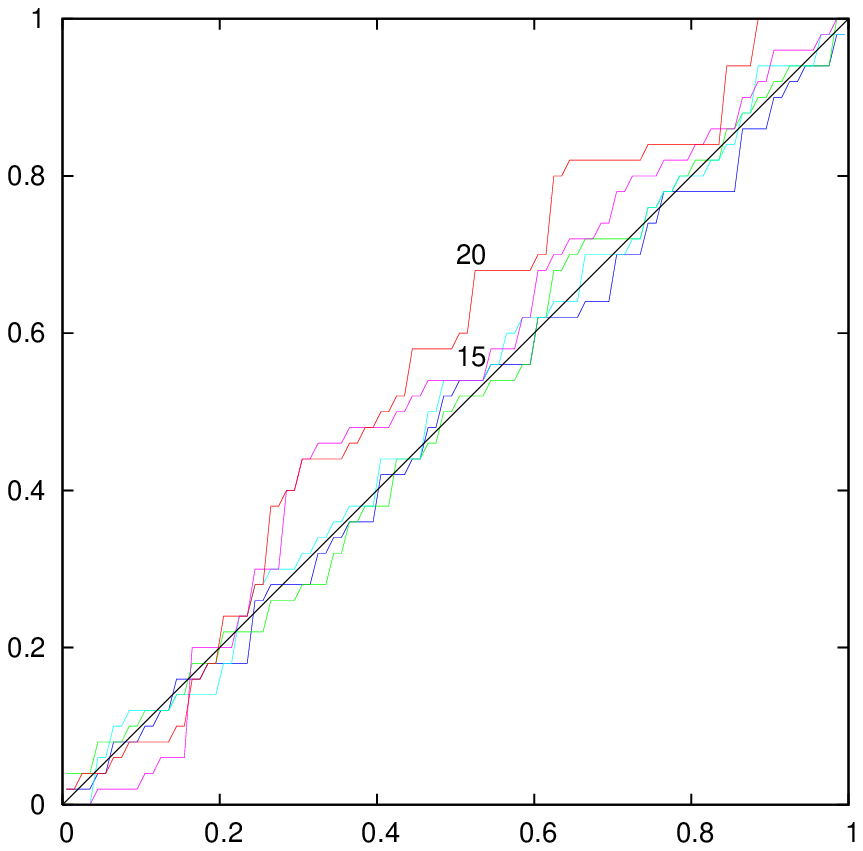}}
\caption{P--P plot of $\kappa$ for $N$=100, with the generating tensor
  values along the x-axis and the inverted tensor values along the
  y-axis. The amount of perturbation in degrees is shown on each
  individual curve.}
  \label{fig:n100}
\end{figure}

The P-P plots obtained from the new analyses presented here (Figure
\ref{fig:n20}--\ref{fig:n100}) show a considerable improvement for the
method compared to the results obtained by
\citet{hardebeck-hauksson01:stress}. It appears that the change from
linear statistics to directional statistics improves the
appropriateness of the confidence interval estimation significantly.
The graphs show a number of trends, including some unexpected ones.

The most obvious, and expected, trend is a decrease in precision of
the confidence interval estimates as the amount of error
increases. This trend is present in each individual graph (Figures
\ref{fig:n20}--\ref{fig:n100}).

What is unexpected is that if the estimate of the confidence intervals
is imprecise, it is systematically too large. In the face of error,
the method performs better than one would expect, something that was
very pronounced in the study by \citet{hardebeck-hauksson01:stress}.

The other noticeable and expected trend is an increase in precision of
the confidence interval estimates as the number of focal mechanisms
increases. This trend is very pronounced in graphs of the stress
tensor perturbation set (Figures \ref{fig:n20}b, \ref{fig:n50}b and
\ref{fig:n100}b). The graphs of the focal mechanism perturbation set
(Figures \ref{fig:n20}a, \ref{fig:n50}a and \ref{fig:n100}a) show an
increase in precision for the smaller amounts of error, but an
unexpected decrease for the larger amounts of error.

There is also a clear tendency for an increased precision for the
stress tensor perturbation sets against their focal mechanism
perturbation equivalents (the (b) subfigure against the (a) subfigure
for each of Figures \ref{fig:n20}, \ref{fig:n50} and \ref{fig:n100}).

\subsection{Mohr Sphere Projections}

\begin{figure}
\subfigure[Generating Stress Tensor]{
\includegraphics[width=120mm,keepaspectratio]{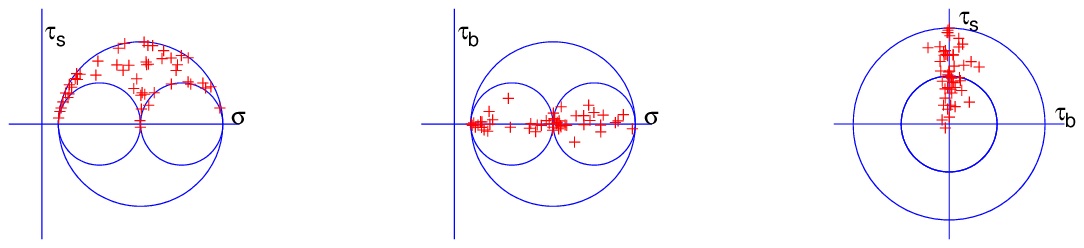}}
\subfigure[Inverted Stress Tensor]{
\includegraphics[width=120mm,keepaspectratio]{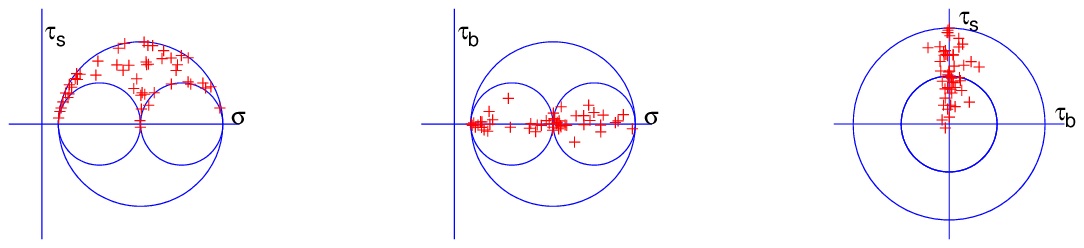}}
\caption{Mohr Sphere projections of the poles of the fault planes
  relative to the stress components defined by the respective stress
  tensor. The fault planes have been subjected to a 1\degree\ Fisher
  distributed perturbation.}
\label{fig:mohr_faults_e01}
\end{figure}

\begin{figure}
\subfigure[Generating Stress Tensor]{
\includegraphics[width=120mm,keepaspectratio]{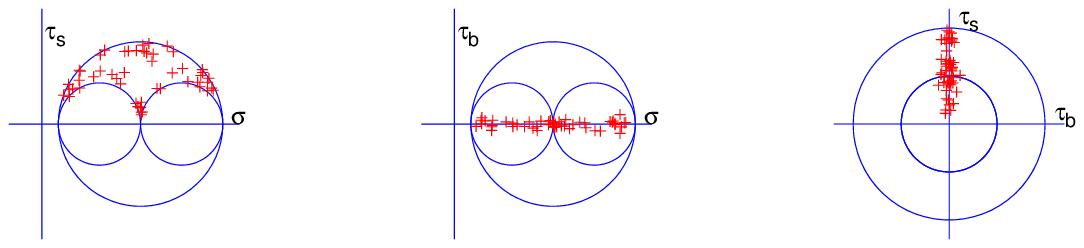}}
\subfigure[Inverted Stress Tensor]{
\includegraphics[width=120mm,keepaspectratio]{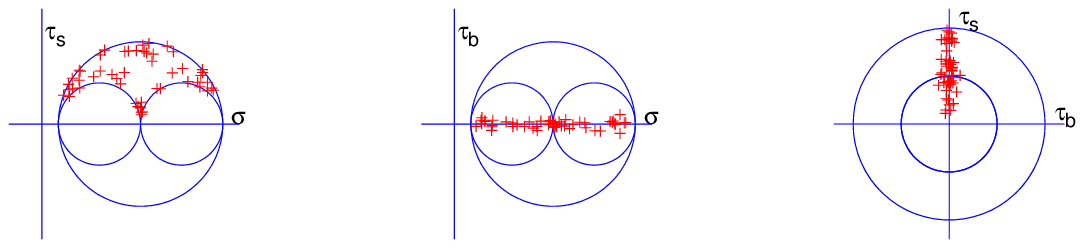}}
\caption{Mohr Sphere projections of the poles of the fault planes
  relative to the stress components defined by the respective stress
  tensor. The stress tensor has been subjected to a 1\degree\ Fisher
  distributed perturbation.}
\label{fig:mohr_tensor_e01}
\end{figure}

\begin{figure}
\subfigure[Generating Stress Tensor]{
\includegraphics[width=120mm,keepaspectratio]{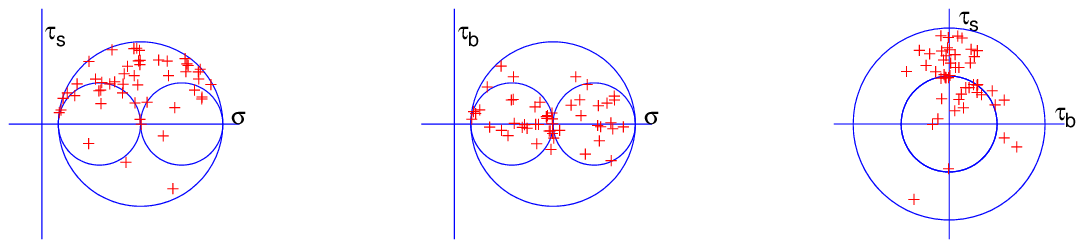}}
\subfigure[Inverted Stress Tensor]{
\includegraphics[width=120mm,keepaspectratio]{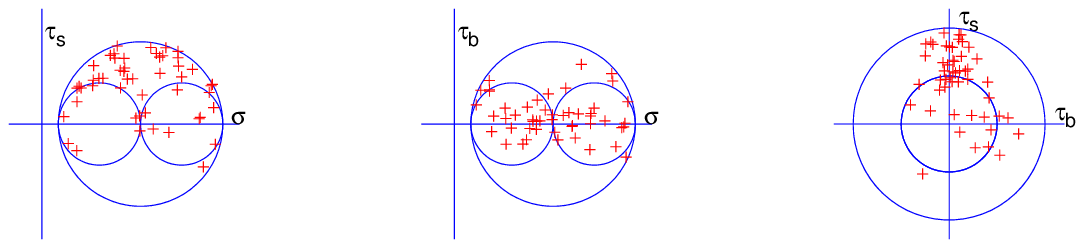}}
\caption{Mohr Sphere projections of the poles of the fault planes
  relative to the stress components defined by the respective stress
  tensor. The fault planes have been subjected to a 15\degree\ Fisher
  distributed perturbation.}
\label{fig:mohr_faults_e15}
\end{figure}

\begin{figure}
\subfigure[Generating Stress Tensor]{
\includegraphics[width=120mm,keepaspectratio]{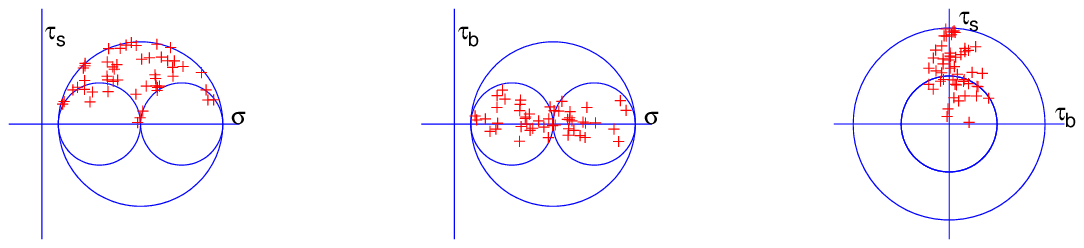}}
\subfigure[Inverted Stress Tensor]{
\includegraphics[width=120mm,keepaspectratio]{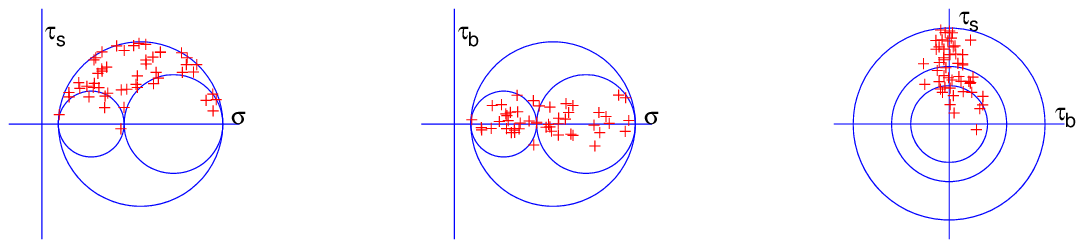}}
\caption{Mohr Sphere projections of the poles of the fault planes
  relative to the stress components defined by the respective stress
  tensor. The stress tensor has been subjected to a 15\degree\ Fisher
  distributed perturbation.}
\label{fig:mohr_tensor_e15}
\end{figure}

The Mohr Sphere projections (Figures
\ref{fig:mohr_faults_e01}--\ref{fig:mohr_tensor_e15}) are examples of
4 data set of 50 focal mechanisms which have undergone either a
1\degree\ (Figures \ref{fig:mohr_faults_e01} and
\ref{fig:mohr_tensor_e01}) or a 15\degree\ (Figures
\ref{fig:mohr_faults_e15} and \ref{fig:mohr_tensor_e15}) Fisher
distributed amount of error. Each figure shows two sets of Mohr Sphere
plots: one illustrating the fault planes against the generating stress
tensor (or its spherical average when it was the generating stress
tensor that underwent perturbation), and one showing the same set of
fault planes against the inverted stress tensor calculated by FMSI.

The first trend which stands out is the larger amount of scatter on
the Mohr Sphere projections for the fault plane perturbed data sets
(Figures \ref{fig:mohr_faults_e01} and \ref{fig:mohr_faults_e15})
compared to the projections for the stress tensor perturbed data sets
(Figures \ref{fig:mohr_tensor_e01} and \ref{fig:mohr_tensor_e15})
which is particularly noticeable on the $\sigma-\tau_b$ graphs. This
increase in scatter includes the presence of fault planes with the
wrong direction of slip ($\tau_S < 0$, seen as crosses which plot
below the X-axis in the $\sigma-\tau_S$ and $\tau_b-\tau_S$ graphs).

A second trend is the larger amount of change between the Mohr Sphere
plots (going from the plot of the fault planes in the generating
stress tensor Mohr Sphere to the Inverted tensor Mohr Sphere) for the
data sets that have undergone fault plane perturbations compared to
the data sets that have undergone stress tensor perturbation. There is
more change visible going from the (a) subfigures to the (b)
subfigures in Figures \ref{fig:mohr_faults_e01} and
\ref{fig:mohr_faults_e15} than in Figures \ref{fig:mohr_tensor_e01}
and \ref{fig:mohr_tensor_e15}.

\section{Discussion}
Theoretical and statistical reasoning requires us to use directional
statistics instead of linear statistics when we deal with focal
mechanisms and deviatoric stress tensor inversion.  The Fisher
distribution and its properties are well established and its use in
the present, theoretical study has been effective. The question does
arise if this distribution is a good representation of actual
seismological data.

Discussions of the distribution of data misfit to stress tensors in
the literature
\citep{gephart-forsyth84:improved,hardebeck-hauksson01:stress,wyss-al92:comparison}
consistently mention the non-normality of the data and refer
qualitatively to exponential or Poisson distributions. Turning to
spherical statistics resolves this issue immediately.  The Fisher
distribution, as given in equation \ref{eq:fisher-distribution}, shows
directly its exponential nature. The method adopted by
\citet{hardebeck-hauksson01:stress} for their simulations, in which
they combine exponentially distributed dip angles with uniformly
distributed azimuth angles, generates in effect a set of Fisher
distributed orientations.

Since we are trying to assess the confidence intervals of the stress
tensor inversion through simulations, we have to ensure that we use
appropriate models and calculations to compare the perturbation
distributions. It is highly instructive to compare the effects of
injecting uncertainty into the simulations at two different points.
It is for this reason that I opted to inject uncertainty or error in
the focal mechanisms to test the inversion process by perturbing the
orientation of the generating stress tensor, as well as the more
intuitive approach of perturbing the orientation of the individual
focal mechanisms directly. 

The fault planes are chosen from uniform spherical random orientations
(equation \ref{eq:random-planes}) and the direction of slip is
calculated from the stress tensor according to equation
\ref{eq:slip-direction}. It is clear that the relation between
orientation of any of the focal mechanisms and the stress tensor is
highly non-linear. That non-linearity necessarily also applies to any
perturbation and in particular to the way in which it propagates in
any calculation or inversion. It is incorrect to assume that a
perturbation distribution applied to a set of focal mechanisms would
translate to the same perturbation distribution of the inverted stress
tensor. This effect can be clarly seen in the Mohr Sphere plots where
the scatter is much larger in the plots of the data sets subjected to
fault plane perturbation (Figures \ref{fig:mohr_faults_e01} and
\ref{fig:mohr_faults_e15}) than those subjected to tensor perturbation
(Figures \ref{fig:mohr_tensor_e01} and \ref{fig:mohr_tensor_e15}).

Comparing the dispersions of the data misfit to the inverted stress
tensor with the dispersion of the perturbations in this case will be
misleading. The dispersion of the stress tensor orientations
(consistent with the perturbed focal mechanisms) is necessarily larger
than the dispersion used perturb the focal mechanisms. It is therefore
not surprising that the inversion method will converge to the correct
stress tensor more often than one would expect from the
dispersion used. This problem can be avoided by letting the perturbation
occur on the generating tensor, before generating the focal
mechanisms. This effect is clearly shown by the contrast between the
P-P plots of the data sets subjected to fault plane perturbation and
the data sets subjected to stress tensor perturbation (respectively
the (a) and (b) subplots in Figures \ref{fig:n20}--\ref{fig:n100}).

Perturbing the fault planes directly in effect introduces bias into
the data set because each of the perturbed fault planes appears to
have been generated by a potentially very different stress tensor,
including incompatibly oriented stress tensors. This bias contaminates
the dispersion. Using this contaminated dispersion to measure the
accuracy of the confidence intervals then leads to the erroneous
conclusion that the inversion method overestimates the confidence
intervals.

As far as simulations are concerned, the statistics can only
legitimately be compared if they are commensurable. That is not the
case when error is injected at the level of the individual fault
plane. Unfortunately, this situation mirrors exactly what happens in
the real world: the parameters describing a calculated focal mechanism
are subject to error and uncertainty, not the generating stress
tensor. Therefore, it may seem that the fact that FMSI yields correct
confidence intervals only when the generating stress tensor is
subjected to error, but fails to do so when focal mechanisms are error
prone, is of no practical benefit.

The paradox can be resolved as follows. I have demonstrated that FMSI
yields correct confidence intervals when the dispersions used to
calculate the confidence intervals are commensurable. I have also
shown that error in focal mechanisms introduces bias in the statistics
of the population at hand. But thanks to the side-by-side simulations,
it is possible to estimate to what extent this bias contaminates the
calculations of FMSI. As one would expect from the law of large
numbers \citep{revesz68:law}, a statistic converges to its theoretical
value as the population size increases. This can be seen very clearly
in the P-P plots: the results of the simulations converge to the
diagonal as the sample size increases. This is very obvious for the
simulations in which the generating stress tensor was subjected to
error (the (b) subplots in Figures
\ref{fig:n20}--\ref{fig:n100}). Closer scrutiny of the P-P plots of
the simulations in which the focal mechanisms were subjected to
perturbation shows a similar, albeit slower convergence (the (a)
subplots in Figures \ref{fig:n20}--\ref{fig:n100}). There is a
trade-off between amount of error and population size: FMSI is capable
of estimating confidence intervals correctly up to a certain amount of
bias. Figure \ref{fig:n20}a shows that for populations of 20 focal
mechanisms, FMSI is capable of dealing with an average of 5\degree\ of
error on the focal mechanisms. This increases to 10\degree\ for 50
focal mechanisms (Figure \ref{fig:n50}a) and to 12--13\degree\ for 100
focal mechanisms (Figure \ref{fig:n100}a).

\begin{figure}
  \includegraphics[width=60mm,keepaspectratio]{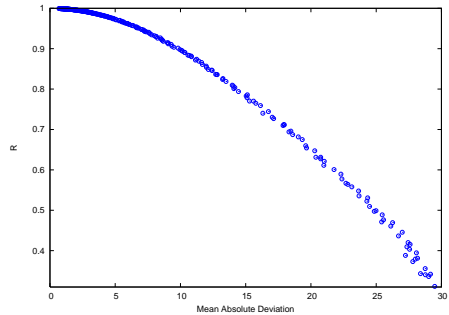}
  \caption{Empirical relation between the $\bar{R}$ values and mean
    absolute deviation (\degree) for the Fisher Distribution}
  \label{fig:r-mad}
\end{figure}

The calculations by FMSI do not use directional statistics and the
questions arises if a simple conversion is possible from the
dispersion measure given by FMSI to directional statistics and allow
the application of the correct calculation of confidence intervals as
shown here.  The mean deviation calculated by FMSI is an average of
angles while the circular mean is the average of the cosines of
angles.  There is no analytical formula to go from one to the
other. Nevertheless, there is a clear relation between the mean
absolute deviation and $\bar{R}$ for the Fisher Distribution (see
Figure \ref{fig:r-mad}) and graphs of this nature could be used to
estimate values which can then be used to determine the confidence
intervals. The correct procedure, albeit more calculation intensive,
is to use the list of individual focal mechanism misfits (which can be
obtained through FMSIETAB) and to apply equation \ref{eq:kappa-chi2}.

\section{Conclusions}
The processing and discussion of data from focal mechanisms and
deviatoric stress tensors should use directional statistics.

The genesis of synthetic data in the context of stress tensor
inversion requires very careful scrutiny of how perturbation is
incorporated.

The inversion of deviatoric stress tensors from focal mechanisms using
the grid search method of Gephart yields reliable confidence intervals
if correct, directional statistics are used. When calculated misfits
are small for moderate to large numbers of focal mechanisms, the
method recovers the true stress tensor and confidence intervals become
superfluous.

Gephart's FMSI is a highly effective and reliable method for stress
tensor inversion.

\section{Data and Resources}
The synthetic data, data inversion and additional calculations relied
on a combination of Bash shell scripts and procedures written for the
Octave program \citep{eaton02:octave}. The FMSI suite of programs is
in the public domain and made available by \citet{gephart90:fmsi}. The
fortran source code is freely available from
www.geo.cornell.edu/pub/FMSI.

 \end{document}